\documentclass{article}
\usepackage{spconf,amsmath,graphicx, amssymb}
\usepackage{float}
\usepackage{cite}
\usepackage{mathtools}

\usepackage{capt-of}
\usepackage{url}
\usepackage[nolist]{acronym}
\usepackage{amssymb,multirow}
\usepackage{booktabs, bm}
\usepackage{booktabs,caption,fixltx2e}
\usepackage{subfigure}
\usepackage{lineno}
\usepackage[]{algorithm2e}
\usepackage{hyperref}
\usepackage{bibspacing}
\usepackage{caption}
\def\RR{{\mathbb R}}

\newacro{MVDR}[MVDR]{Minimum Variance Distortionless Response}
\newacro{GEV}[GEV]{Generalized Eigenvalue Decomposition}
\newacro{MCWF}[MCWF]{Multi-Channel Wiener Filter}
\newacro{SDW-MWF}[SDW-MWF]{Speech Distortion Weighted MWF}
\newacro{SISDR}[SI-SDR]{Scale Invariant Signal-to-Distortion Ratio}
\newacro{STFT}[STFT]{Short-Time Fourier Transform}
\newacro{DNN}[DNN]{Deep Neural Network}
\newacro{SNR}[SNR]{Signal-to-Noise-Ratio}
\newacro{FLOPs}[FLOPs]{Floating Point Operations}
\newacro{SIR}[SIR]{Signal-to-Interference Ratio}
\newacro{ASR}[ASR]{Automatic Speech Recognition}
\newacro{IBM}[IBM]{Ideal Binary Mask}
\newacro{IRM}[IRM]{Ideal Ratio Mask}
\newacro{WLM}[WLM]{Wiener-Like Mask}
\newacro{SDR}[SDR]{Signal-to-Distortion Ratio}
\newacro{STOI}[STOI]{Short-time Objective Intelligibility}
\newacro{MSE}[MSE]{Mean-Squared Error}
\newacro{WER}[WER]{Word Error Rate}
\newacro{WPE}[WPE]{Weighted Prediction Error}
\newacro{MISO}[MISO]{Multiple Input Single Output}
\newacro{DFL}[DFL]{Deep Feature Loss}

\title{Multi-Channel Target Speaker Extraction with Refinement:\\The WavLab Submission to the Second Clarity Enhancement Challenge}

\name{
\begin{tabular}{c}
\it Samuele Cornell$^{1,2}$, Zhong-Qiu Wang$^2$, Yoshiki Masuyama$^{3, 2}$,  Shinji Watanabe$^2$ \\ Manuel Pariente$^{4}$, 
Nobutaka Ono$^{3}$
\end{tabular}
}
\address{
$^1$Università Politecnica delle Marche, Italy\,\,
$^2$Carnegie Mellon University, USA\\
$^3$Tokyo Metropolitan University, Japan\,\,
$^4$Pulse Audition, France\\
{\small\texttt{\{cornellsamuele, wang.zhongqiu41\}@gmail.com}}
}

\begin{document}
\ninept
\maketitle

\setlength{\abovedisplayskip}{3.5pt}
\setlength{\belowdisplayskip}{3.5pt}

\begin{abstract}

\noindent This paper describes our submission to the Second Clarity Enhancement Challenge (CEC2), which consists of target speech enhancement for hearing-aid (HA) devices in noisy-reverberant environments with multiple interferers such as music and competing speakers.
Our approach builds upon the powerful iterative neural/beamforming enhancement (iNeuBe) framework introduced in our recent work, and this paper extends it for target speaker extraction.
We therefore name the proposed approach as iNeuBe-X, where the ``X'' stands for extraction.
To address the 
challenges encountered in the CEC2 setting, we introduce four major novelties:  
(1) we extend the state-of-the-art TF-GridNet model \cite{zqinprep}, originally designed for monaural speaker separation, for multi-channel, causal speech enhancement, and large improvements are observed by replacing the TCNDenseNet used in iNeuBe \cite{lu2022towards} with this new architecture;  
(2) we leverage a recent dual window size approach with future-frame prediction \cite{wang2022stft} to ensure that iNueBe-X satisfies the $5$ ms constraint on algorithmic latency required by CEC2;
(3) we introduce a novel speaker-conditioning branch for TF-GridNet to achieve target speaker extraction;
and (4) we propose a fine-tuning step, where we compute an additional loss with respect to the target speaker signal compensated with the listener audiogram.
Without using external data, on the official development set our best model reaches a hearing-aid speech perception index (HASPI) score of $0.942$ and a scale-invariant signal-to-distortion ratio improvement (SI-SDRi) of $18.8$\,dB.
These results are promising given the fact that the CEC2 data is extremely challenging (e.g., on the development set the mixture SI-SDR is $-12.3$\,dB).
A demo of our submitted system is available at \href{https://popcornell.github.io/WAVLab_CEC2_demo/}{WAVLab\_CEC2\_demo}. 
\end{abstract}
%


\vspace{-0.2cm}
\section{Proposed Methods}

\subsection{iNeuBe-X System Overview}\label{sec:ineube-x}

\begin{figure}
  \centering  
  \includegraphics[width=8.5cm]{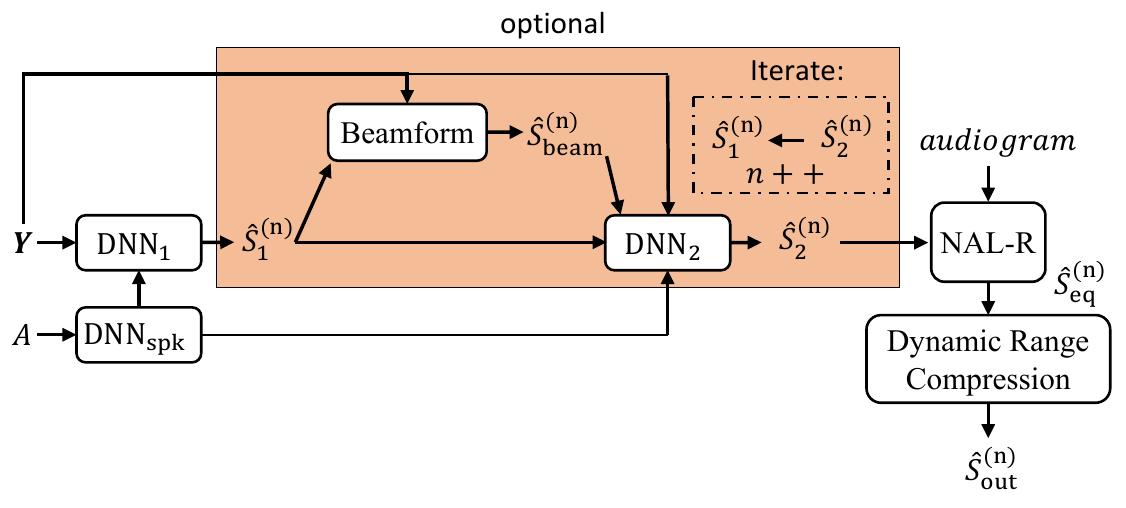} \vspace{-0.2cm} \\  
  \caption{Overview of proposed iNeuBe-X framework.
  We employ a causal multi-channel Wiener filter beamformer between DNN$_1$ and DNN$_2$.} 
  \label{fig:ineubex}
  \vspace{-0.5cm}
\end{figure}

Our proposed iNeuBe-X system is depicted in Fig.~\ref{fig:ineubex}.
It is motivated by the iNeuBe framework~\cite{lu2022towards}, which contains two multi-microphone input single-microphone output (MISO) DNNs~\cite{Wang2020css}, with the first one producing an initial target estimate for beamforming and the second one performing post-filtering for better enhancement.
One key difference is that here we condition each MISO DNN with a speaker-enrollment embedding extracted from a third DNN$_\text{spk}$ module, which is jointly trained with the rest and takes as input an enrollment utterance of the target speaker.
Given the complex multi-channel short-time Fourier transform (STFT) spectrogram of the mixture signal, $\bm{Y}$, the first DNN (denoted as DNN$_1$ in Fig.~\ref{fig:ineubex}) estimates the complex STFT spectrogram $\hat{S}_1^{(n)}$ (at iteration $n=1$) of the anechoic target speech captured at a reference HA array mic, which is set to ``CH1 Left'' in CEC2.
This estimate can then be refined by a second step, which involves a DNN-supported beamformer, and a second MISO DNN (denoted as DNN$_2$) with additional inputs from the outputs of DNN$_1$ and the beamformer.
This refinement can be iteratively performed, following \cite{lu2022towards}.
Our system is trained to operate with a $32$\,kHz sampling rate and with an STFT with $16$\,ms window, $4$\,ms stride, and the square-root Hann window.
This would give such system a theoretical latency of $16$\,ms~\cite{wang2022stft}.
To overcome this, we train both DNN$_1$ and DNN$_2$ to predict the current plus future $3$ frames of the target anechoic signal, so that a new synthesized frame output can be obtained for each $4$\,ms stride.
This is done in practice by padding zeros to the left of the input mixture STFT. 
The beamforming operation is performed in a frame-online fashion to comply with the 5\,ms constraint on algorithmic latency required in CEC2.
We employ a causal multi-channel Wiener filter as in \cite{lu2022towards, wang2022stft}.
Since the listener head can rotate in CEC2, instead of simply accumulating the statistics, we use the recursive averaging strategy~\cite{Gannot2017} with a $0.5$ forgetting factor.

\vspace{-0.2cm}
\subsection{Speaker Enrollment Embedding Extraction}

For DNN$_\text{spk}$, we use the encoder and TCN modules (and remove the decoder) of the TCNDenseNet architecture \cite{Wang2021LowDistortion, lu2022towards} to extract speaker embeddings.
As the enrollment utterances are available beforehand, this DNN can be non-causal and its output can be cached before inference for each target speaker.
Given a complex spectra $A$ of a monaural adaptation utterance, this DNN extracts a fixed-length speaker embedding $X_\text{adapt} \in \RR^{128}$ by performing mean-pooling over time on the outputs of the TCN module.
The speaker embedding is then leveraged as an additional cue by the iNeuBe framework for target extraction.
Differently from \cite{Wang2021LowDistortion, lu2022towards}, we use a smaller DNN, with 3 TCN repeats, 4 TCN blocks, 128 TCN channels, and 16 hidden channels in the down-sampling Conv2D and DenseNet layers of the encoder. 
This amounts to a total of around $0.6$\,million (M) parameters. 



\vspace{-0.2cm}
\subsection{Frame-online Speaker-conditioned MISO-TF-GridNet}\label{sec:mc_csm}

\begin{figure}
  \centering  
  \includegraphics[width=5cm]{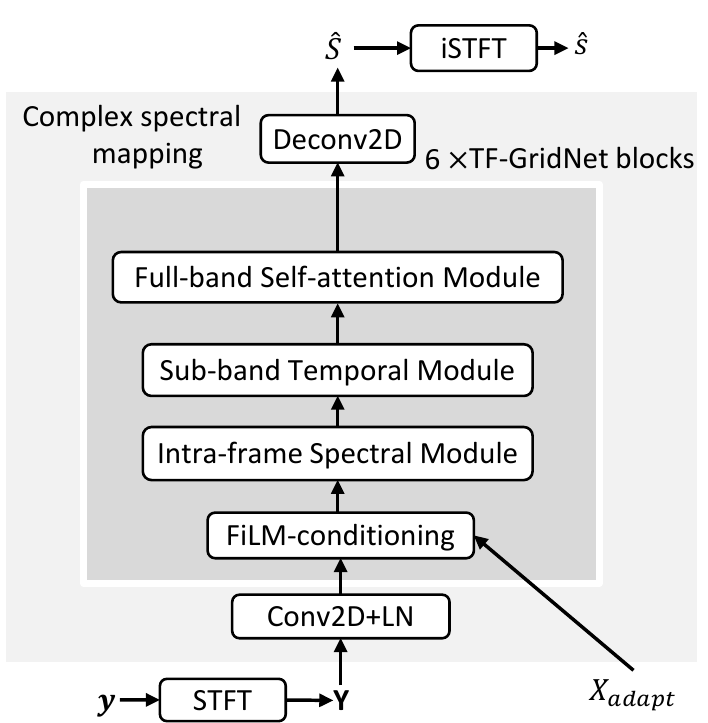}
  \vspace{-0.1cm}
  \caption{Proposed frame-online speaker-conditioned MISO-TF-GridNet.}
  \label{fig:tf_mpdnn_overview}
  \vspace{-0.4cm}
\end{figure}

For both DNN$_1$ and DNN$_2$ we use TF-GridNet~\cite{zqinprep}, which is a state-of-the-art separation model that recently achieved an impressive 23.4\,dB SI-SDRi on the WSJ0-2mix dataset~\cite{Hershey2016DC}.
Motivated by this strong result, we extended this model to multi-channel scenarios as depicted in Fig.~\ref{fig:tf_mpdnn_overview}.
Since it uses complex spectral mapping, we feed all channels by simply stacking the 
real and imaginary (RI) components of all input channels \cite{lu2022towards} to TF-GridNet, converting it to a MISO model named as MISO-TF-GridNet.
We use feature-wise linear modulation (FiLM) \cite{perez2018film} over the speaker embedding extracted by DNN$_\text{spk}$ at the beginning of each TF-GridNet block to condition the separation. 
Note that TF-GridNet proposed in \cite{zqinprep} is non-causal. 
Here we make the following modifications to achieve frame-online processing. 
Firstly, we use layer normalization (LN) after the first Conv2D layer.
Secondly, we use a causal full-band self-attention module by masking the attention matrix.
Thirdly, we use a modified sub-band temporal module, where we employ an unidirectional LSTM (instead of bidirectional as in \cite{zqinprep}) and, in the unfolding operation, we stack only the previous frames' features with the current.
This way, the sub-band temporal module can output one frame for each frame in the input if the stride is set to $1$. 

We use the same hyper-parameters suggested in \cite{zqinprep}, except that $E$ is set to $2$ here. 
The resulting TF-GridNet has around $8$\,M parameters and the full iNeube-X model (with DNN$_1$ and DNN$_2$) roughly doubles that amount. 

\vspace{-0.3cm}


\subsection{STFT-domain NAL-R and Dynamic Range Compression}

We found that enhancement alone is not sufficient for obtaining an high HASPI. Without any compensation to listener's hearing loss, the HASPI for the oracle target speech is less than $0.7$ on the development set.
To perform fitting in a frame-online fashion, we perform the NAL-R fitting \cite{byrne1986national} and dynamic range compression in the STFT domain.
We followed the dynamic range compression algorithm implemented in \cite{mccormack2017fft}, and set the knee to -10\,dB, threshold -40\,dB, ratio 1.2, knee width 4, attack 0.05 s, and release 0.2\,s, respectively.
As we use in iNeuBe-X 16\,ms STFT windows, the length of the NAL-R equalization filter was set to 80 taps. 
When fed the oracle target anechoic speech, our STFT-domain NAL-R fitting plus compression algorithms produces an HASPI score of $0.987$, which is slightly lower than the $0.998$ score obtained by the CEC2 baseline fitting algorithm.
This is likely due to the shorter NAL-R filter used in our algorithms, but our oracle score is still acceptable. 

\subsection{System Training Configurations and External Data Usage}

We trained models on two datasets, the original CEC2 data, which we scaled up to $16$k scenes using the provided scene generation script, and this $16$k scenes plus additional $6$k scenes generated using the same configuration except for external clean speech taken from LibriVox in order to increase speaker variations. 

Regarding training, we first train DNN$_1$ with DNN$_\text{spk}$ to convergence, and then freeze these two and train DNN$_2$ using the output of these two plus the MCWF ouput. 
As explained in Section \ref{sec:ineube-x}, the two MISO networks are trained to estimate the target anechoic signal at CH1 Left from the multi-channel mixture signal using complex spectral mapping.  
We use the Adam optimizer with $0.001$ learning rate and batch size $1$. We clip gradients with $L_2$ norm more than $1$.
The MISO DNNs are trained on random $4$\,s segments of the input signal, which are constrained in such a way that the target and the interferer-only beginning part should appear both at least for $1$\,s.
We use the whole development set for validation. 
The learning rate is halved if the validation loss is not improved over 5 epochs.
We follow the scale-invariant loss function suggested in~\cite{lu2022towards}, which is defined on the estimated time-domain signal and its magnitude.
Differently from \cite{lu2022towards}, we use a multi-resolution STFT filterbanks with $512$-, $1024$-, $2048$-, $256$- and $128$-sample windows to compute multiple magnitude losses.
We observed that the magnitude loss leads to clearly better HASPI.
Since this loss is scale-invariant, to prevent clipping in inference we re-scale the MISO DNNs' estimates in a frame-online fashion using the output of the MCWF. 

To further improve HASPI, after training DNN$_2$ to convergence, we fine-tune DNN$_2$ by adding an additional loss term where the same multi-resolution loss is computed 
between NAL-R fitted target anechoic speech and NAL-R fitted DNN estimates.
Since listener audiograms are not available in the training set, we randomly sample audiograms from the development set listeners to perform this fitting-aware training. 


\begin{table}[]
\scriptsize
\centering
\caption{Results on development set of CEC2.
}
\vspace{-0.2cm}
\label{tab:dev_res}
\setlength{\tabcolsep}{2.5pt}
\begin{tabular}{lcccc}
\toprule

Approaches & SI-SDRi (dB) & STOI & PESQ & HASPI \\
 \midrule
Challenge baseline (mixture CH1 Left) \cite{ren2021neural} & 0.0  & 0.563   & 1.197 & 0.245 \\
Challenge baseline (oracle CH1 Left) & $\infty$  & 1.0  & 5.0  & 0.998 \\
\midrule
BLSTM-WPE+MVDR~\cite{lu2022espnet} & 10.481 & 0.743 & 1.931 & 0.436 \\
iNeuBe DNN$_1$~\cite{lu2022towards} & 16.567 & 0.887  &  1.946  & 0.723 \\
\midrule
iNeuBe-X DNN$_1$ & 17.97  & 0.92 & 1.951  &  0.842 \\ 
\quad+ DNN$_2$ + NAL-R fine-tune & 19.082 & 0.947 & 2.269  & 0.942 \\
\quad+ DNN$_2$ + NAL-R fine-tune + External Data & 19.514  & 0.954 & 2.336 & 0.943 \\  
\bottomrule
\end{tabular}
\vspace{-0.4cm}
\end{table}
\vspace{-0.4cm}
\section{Results on Development Set}\label{sec:results}
\vspace{-0.2cm}
We report our results obtained on the CEC2 development set in Table~\ref{tab:dev_res} in terms of SI-SDR, STOI, PESQ and HASPI.
Note that SI-SDR, STOI and PESQ are computed before listener adaptation and compression while HASPI is computed after.
Together with the proposed iNeuBe-X, we also report the performance for two non-causal methods: iNeube (only DNN$_1$, based on TCNDenseNet)~\cite{lu2022towards} and a DNN-based masked beamforming approach from the ESPNet-SE++ toolkit~\cite{lu2022espnet}, which employs a two-layer bidirectional LSTM (BLSTM) and weighted prediction error (WPE) followed by minimum variance distortionless response (MVDR) beamforming. 
As another comparison, we report the scores of the unprocessed mixtures and the oracle target speech.
The latter can be considered as a reasonable upper bound for our proposed system, since we employ a similar compensation strategy. 
Note that the CEC2 baseline system does not perform enhancement, and only performs listener adaptation and dynamic range compression. 
For our proposed systems, we report performance for various configurations and training strategies in the third panel: iNeube-X with only DNN$_1$,
iNeube-X with DNN$_2$ trained using the additional loss after listener adaptation (+ DNN$_2$ + NAL-R fine-tune), and with external data (+ DNN$_2$ + NAL-R fine-tune + External Data).
The two latter systems are the ones we submitted to CEC2. We found that an additional iteration of beamforming followed by post-filtering led to slightly better HASPI, at a cost of increased computational overhead.
Thus we only submitted signals obtained after the first DNN$_2$ refinement step.

We emphasize that our system enjoys an algorithmic latency of 4 ms, and we confirm that the proposed system satisfies the $5$\,ms constraint on algorithmic latency using this public code\footnote{https://github.com/zqwang7/CausalityCheck}.

\vspace{-0.2cm}
\section{Acknowledgements}
\vspace{-0.2cm}

S. Cornell was partially supported by Marche Region within the funded project ``Miracle'' POR MARCHE FESR 2014-2020.
Z.-Q. Wang used the Extreme Science and Engineering Discovery Environment ~\cite{xsede}, supported by the NSF under grant ACI-1548562, and the Bridges system~\cite{nystrom2015bridges}, supported by the NSF under grant ACI-1445606, at the Pittsburgh Supercomputing Center.
Y. Masuyama and N. Ono were partially supported by JSPS KAKENHI Grant Numbers JP21J21371 and JST CREST Grant Number JPMJCR19A3.

\bibliographystyle{IEEEtran}
{\footnotesize
\bibliography{refs}
}

\end{document}